\documentclass[10pt,a4paper]{article}

\usepackage[T1]{fontenc}
\usepackage{epsf} 
\usepackage{amsmath}
\usepackage{amssymb}
\usepackage{graphicx}
\usepackage{booktabs} 
\usepackage[small]{caption}
\usepackage{multirow}

\newcommand{\scar}[1]{``#1''}

\newcommand{\nlsm}[1]{nonlinear sigma model#1}{}
{}
{}

\newcommand{\ie}{i.e.\,}

\newcommand{\prodl}{\prod\limits}
\newcommand{\suml}{\sum\limits}
\newcommand{\intl}{\int\limits}

\newcommand{\dm}{\mathrm{d}}

\newcommand{\Aeff}{\mathcal{A}_{\mathrm{eff}}}

\newcommand{\bpi}{\boldsymbol{\pi}}

\newcommand{\cov}[1]{\partial_{#1}}

\newcommand{\op}[1]{\mathbf{#1}}

\newcommand{\Dstar}[1]{D^{*}\!(#1)}

\newcommand{\Gbars}{\bar{G}^{*}}

\newcommand{\ddGbars}{\partial_0^2\bar{G}^{*}}

\newcommand{\ededt}{\dot{\vec{e}}(t)\dot{\vec{e}}(t)}

\newcommand{\delN}{\delta}
\newcommand{\delM}{r}

\newcommand{\ket}[1]{|#1\rangle }

\newcommand{\norm}[1]{\langle #1 | #1 \rangle}
\newcommand{\expv}[3]{\langle #1 | #2 | #3\rangle}

\newcommand{\egy}[2]{\varepsilon^{(#1)}_{#2}}

\newcommand{\gr}[2]{#1(#2)}
\newcommand{\ordo}[1]{\mathcal{O}(#1)}

\newcommand{\fm}{\,{\rm fm}}
\newcommand{\MeV}{\,{\rm MeV}}

\begin{document}

\begin{titlepage}

  \phantom{}
  \vskip 3 true cm

  \begin{center}
    {\Large\bf 
      The QCD rotator with a light quark mass}
  \end{center}
  \vskip 1 true cm
  \centerline{\large Manuel Weingart}
  \vskip1ex
  \centerline{\it Albert Einstein Center for Fundamental Physics,}
  \centerline{\it Institute for Theoretical Physics, University of Bern}
  \centerline{\it Sidlerstrasse 5, CH-3012 Bern, Switzerland}
  \vskip 1 true cm
  	\centerline{\bf Abstract}
	The low-lying energy excitations of 2-flavour QCD in the chiral limit confined to a small spatial box ($\delta$-regime) are that of an $\gr{O}{4}$ rotator. In this work, we calculate the corrections due to the presence of a 		nonzero quark mass up to NNL order by means of dimensional regularised chiral
 	perturbation theory. The final result for the energy gap of the system
 	only involves  the low-energy constants $F$, $\Lambda_1$, $\Lambda_2$ and $B$.

  \vfill
  \eject

\end{titlepage}

%
\section{Introduction and summary}

Quantum chromodynamics, the quantum field theory describing the interactions between  quarks and gluons, is non-perturbative at low energies (large distances). An expansion in the coupling constant breaks down, since the latter becomes large (of $\ordo{1}$) at low energies. In order to study QCD properties at low energies, we have to apply other methods. 

The lattice provides the only known non-perturbative treatment of QCD in terms of the fundamental fields, the quarks and the gluons. However, lattice QCD
calculations are very time-consuming, and so we are restricted to relatively
small volumes or relatively large lattice spacings. Thus, the results we obtain from lattice QCD simulations suffer from finite size effects and discretisation errors. 

In this work, we will restrict ourselves to two light quark flavours in the isospin limit ($m_u=m_d=m$). QCD is known to be spontaneously broken in case the $u$ and $d$ quarks are massless. The QCD Lagrangian is invariant under the global $\gr{SU}{2}_L\times \gr{SU}{2}_R$ transformations, the vacuum of QCD is only invariant under global $\gr{SU}{2}_V$ transformations. According to Goldstone's theorem~\cite{Goldstone:1961eq}, the breaking of a global, continuous symmetry gives rise to massless particles, the Goldstone bosons (GB). 

At low energies, QCD can be described by chiral perturbation theory (ChPT) \cite{Weinberg:1979a96,Gasser:1983yg,Gasser:1983kx}, an effective field theory with pions as degrees of freedom. The low-energy coefficients (LECs), associated to the operators, showing up in effective Lagrangian, have to be determined by experiments or by measurements from lattice simulations. Since ChPT can also be applied if the system is enclosed in a finite volume~\cite{Gasser:1986vb,Gasser:1987zq}, finite volume effects can be calculated analytically.

We consider a volume $V=L_s\times L_s\times L_s\times L_t$, where the temporal extent of the box and the Compton wavelength of the pion are both much larger than the spatial size of the box, $L_t\gg L_s$,  $ML_s\ll 1$ respectively. This special regime is called $\delta$-regime ($\delta$-expansion)~\cite{Leutwyler:1987ak}. Here, $M$ refers to the leading term for the pion mass in infinite volume
\begin{equation}\label{eq: pion mass at infinite volume}
 M^2 = 2mB\,.
\end{equation}
Since $\gr{SU}{2}_L\times\gr{SU}{2}_R$ is the covering group of $\gr{O}{4}$, we use a $\gr{O}{4}$ \nlsm{} in $d=3+1$ Euclidean space-time dimensions to describe the effective theory.

Due to the large Compton wavelength of the pion, large compared to the spatial extent of the box, the system exhibits a global mode which slowly rotates in the internal space. These slow modes and the related energy excitations interest us in this work.  In the chiral limit ($m\to 0$), the low-lying energy excitations of this system can be 
described by the spectrum of the $\gr{O}{4}$ rotator
\begin{equation}\label{eq: O(4) rotator spectrum}
E_\ell = \frac{\ell(\ell+2)}{2\Theta}\,,\qquad \ell=0,1,\ldots\,,
\end{equation}
where this form is only valid up to NNL order. Here, $\ell$ denotes the $\gr{O}{4}$ \scar{angular momentum} (isospin) and $\Theta$ the moment of inertia which is given as
$\Theta=F^2L_s^3$ at leading order~\cite{Leutwyler:1987ak,Fisher:1985zz,Brezin:1985xx}.
The low energy constant $F$ is the pion decay constant in
the chiral limit.  Allowing for higher order terms in the effective
Lagrangian the moment of inertia will receive fast mode corrections proportional to
$1/(FL_s)^2$  at NL order~\cite{Hasenfratz:1993vf}, 
respectively proportional to $1/(FL_s)^4$ at NNL order~\cite{Hasenfratz:2009mp}. We will denote the dimensionless expansion parameter as
\begin{equation}\label{eq: delta0}
\delN^2 = \frac{1}{F^2L_s^2}
\end{equation}
throughout this work.

For sufficiently small quark masses $M\leq 1/F^2L_s^3$, the system is still dominated by the rotator spectrum. The latter then obtains small corrections due to the symmetry breaking terms. The dimensionless expansion parameter related to the symmetry corrections will be denoted by
\begin{equation}
\delM = F^2L_s^3M \,.\label{eq: deltaM}
\end{equation}
The NLO correction to the rotator spectrum coming from the symmetry breaking terms is proportional to $r^4$~\cite{Leutwyler:1987ak}. 

In Section~\ref{sec: separate
  slow and fast modes}, we will illustrate how to separate the slow modes from the fast modes. In order to apply perturbation theory, the effective action has to be expanded in terms of the fast modes. A detailed discussion for the
symmetry-breaking terms is given in Section~\ref{sec: expand
  symmetry breaking action up to NNL}. In Section~\ref{sec: integrate out fast modes}, we will integrate out the fast modes. The calculations are performed in the zero temperature limit
\mbox{($L_t\to\infty$)}, and we use dimensional regularisation (DR). This setup
is in accordance with calculations in~\cite{Hasenfratz:2009mp}. 

The remaining quantum mechanical $\gr{O}{4}$ rotator, now in a external \scar{magnetic} field, is characterised by the moment of inertia, and  by the symmetry-breaking, external field $\eta$, where $\Theta$ and $\eta$ contain corrections from the fast modes.

Finally, we apply perturbation theory in quantum mechanics in order to determine the corrections to the unperturbed rotator spectrum.  
Since the two expansion parameters enter at NLO as $\delN^2$,  and $\delM^4$ respectively, we consider these two combinations to be of the same order $\delN^2\sim\delM^4=\ordo{\delta^2}$. We are interested in corrections up to NNLO, therefore we have to take into account any terms up to  $\ordo{\delta^4}$.

We quote the final result for the energy gap $E_{L_s}$ up to NNL order
\begin{equation}\label{eq: energy gap final result}
E_{L_s}  = \frac{3}{2\Theta}\left[1+\frac{(\Theta\eta)^2}{15}-\frac{193}{120}\frac{(\Theta\eta)^4}{15^2}\right]
\end{equation}
where $\Theta$ is given in~\cite{Hasenfratz:2009mp}  as
\begin{equation}\label{eq: moment of inertia up to NNL}
\begin{split}
\Theta & = F^2 L_s^3\Bigg[1-\frac{2 \Gbars}{F^2L_s^2} + \frac{1}{F^4 L_s^4}\bigg[0.088431628 \\
	& \phantom{=F^2L_s^3\Bigg[} + \ddGbars\frac{1}{3\pi^2}\Big(\frac{1}{4}\log(\Lambda_1L_s)^2+\log(\Lambda_2L_s)^2\Big)\bigg]\Bigg]\,,
\end{split}
\end{equation}
and 
\begin{equation}\label{eq: eta incl NNLO corrections}
\eta =F^2L_s^3M^2\bigg(1-\frac{3\Gbars}{2 F^2L_s^2}\bigg)\,.
\end{equation}
$\Gbars$ and $\ddGbars$ are given by the following numbers
\begin{align}
 \Gbars & = -0.2257849591 \label{eq: value of gbarstar}\\
 \ddGbars & =  -0.8375369106 \label{eq: value of ddgbarstar}\,.
\end{align}
The low-energy constants $l_1$ and $l_2$ enter~\eqref{eq: moment of inertia up to NNL} over their intrinsic scales $\Lambda_1$, respectively $\Lambda_2$~\cite{Colangelo:2001df}.


\section{The effective action up to $\ordo{p^4}$}
The low-energy properties of QCD with two nearly massless flavours, considered in infinite volume, can be covered
by an effective field theory in terms of the pseudo-Goldstone bosons, the
pions. We choose to describe the GB dynamics by an $\gr{O}{4}$ \nlsm{}. The
effective Lagrangian is ordered in a systematic way by the
increasing number of derivatives and increasing power of the symmetry-breaking
field $\vec H$. Derivatives are counted as $\ordo{p}$ and the symmetry-breaking field is counted as $\ordo{p^2}$. Using the convention which identifies the term
$\mathcal{A}_{(d,h)}$ in the effective action containing operators with $d$ derivatives and the field $\vec H$ to the power $h$, we write
\begin{equation*}\label{eq: A_eff up to O(p4)}
\mathcal{A}_{\mathrm{eff}} = \mathcal{A}_{(2,0)}+\mathcal{A}_{(0,1)}+\mathcal{A}_{(4,0)}+\mathcal{A}_{(2,1)}+\mathcal{A}_{(0,2)}+\ldots\,,  
\end{equation*}
for the effective action up to $\ordo{p^4}$. The explicit expressions for these terms are 	
\begin{align}
\mathcal{A}_{(2,0)} & = \phantom{-}\int\!\!\dm x\, \frac{F^2}{2}\cov{\mu}\vec S(x)\cov{\mu}\vec S(x)\,, \label{eq: A20}\\
\mathcal{A}_{(0,1)} & = - \int\!\!\dm x\, \Sigma\vec H\vec S(x)\,,\label{eq: A01}\\
\mathcal{A}_{(4,0)} & = \phantom{-}\int\!\!\dm x\, \frac{1}{4}g_4^{(2)}\left(\cov{\mu}\vec S(x)\cov{\mu}\vec S(x) \right)^2  + \frac{1}{4}g_4^{(3)}\left(\cov{\mu}\vec S(x)\cov{\nu}\vec S(x)\right)^2\,,\label{eq: A40}\\
\mathcal{A}_{(2,1)} & = \phantom{-}\int\!\!\dm x\, k_1\frac{\Sigma}{F^2}\left(\vec H\vec S(x)\right)\left(\cov{\mu}\vec S(x)\cov{\mu}\vec S(x)\right)\,,\label{eq: A21}\\
\mathcal{A}_{(0,2)} & = -\int\!\!\dm x\, k_2\frac{\Sigma^2}{F^4}\left(\vec H\vec S(x)\right)^2\,.\label{eq: A02}
\end{align}
$\vec S(x)$ is a $4$-component vector of unit length,  $\vec S(x)^2=1$, and we choose the external field $\vec H$ to point in the zeroth direction $\vec H=(H,0,...,0)$. Terms which do not depend on $x$ at all have been omitted, since they will enter as an overall factor in the path integral representation. We have used the same conventions as in~\cite{Hasenfratz:1989pk} for the low-energy constants. 

Due to the explicit symmetry breaking, the pions acquire a mass. If we consider the external field $H$ to be small, the pseudo-Goldstone boson mass is given by~\cite{Hasenfratz:1989pk} 
\begin{equation}\label{eq: pion mass for small external field H}
 M^2 = \frac{\Sigma H}{F^2}\,,
\end{equation}
at leading order in $H$. In the following, we will replace the combination
$\Sigma H$ by $M^2 F^2$.  By comparing Eq.~\eqref{eq: pion mass for small
  external field H} with Eq.~\eqref{eq: pion mass at infinite
  volume}, we can identify the external symmetry breaking parameter $H$ with
the quark mass $m$ and $\Sigma$ with $BF^2$.

The effective action~\eqref{eq: A20}-\eqref{eq: A02} is based on conventions from condensed matter physics. In chiral perturbation theory, the pion fields are usually parametrised by $\gr{SU}{2}$ matrices, and the low-energy constants in the $\ordo{p^4}$ effective Lagrangian are labelled by $l_1$, $l_2$, $l_3$, $l_4$ and $h_1$. We determine the relations between the LECs by comparing the $\ordo{p^4}$ effective action in~\cite{Gerber:1988tt} with the terms~\eqref{eq: A20}-\eqref{eq: A02}
\begin{align*}
 g_4^{(2)} &= -4l_1\,, 	&	g_4^{(3)} &=-4l_2\,, 	&\\
 k_1& = l_4\,,		&	k_2 & = l_3 + l_4\,, 	& k_3 & = h_1\,.
\end{align*}
%


\section{Separating the fast modes from the slow modes}\label{sec: separate slow and fast modes}

Chiral perturbation theory can also be applied in finite volume~\cite{Gasser:1986vb,Gasser:1987zq}. In the $\delta$-regime ($L_t\gg L_s$ and $ML_s\ll1$), the fields $\vec S(x)$on a given time slice are strongly correlated  and exhibit a net \scar{magnetisation}

\begin{equation}\label{eq: net magnetisation delta-regime}
 \vec m(t) = \frac{1}{V_s}\int\!\!\dm \vec{x}\, \vec{S}(t,\vec{x})\,,\qquad \vec m(t) = m(t) \vec e(t)\,.
\end{equation}
The direction $\vec e(t)$ of the net "magnetisation" performs a slow rotation in the internal space. These slow modes have to be treated non-perturbatively. The fluctuations (fast modes) around direction of the \scar{magnetisation} can be integrated out in perturbation theory. Therefore, the fast modes have to be separated from the slow modes.

 We incorporate the collective behaviour of the variables $\vec S$ by introducing 
\begin{equation*}
 1 = \prodl_t\int\!\!\dm \vec m(t)\delta^{(N)}\bigg[\vec m(t) -\frac{1}{V_s}\int\!\!\dm\vec x\, \vec S(t,\vec x)\bigg]
\end{equation*}
into the partition function of the system. The partition function then reads as
\begin{equation}
 \begin{split}
  Z = & \prodl_x \int\!\!\dm \vec S(x)\,\delta\bigg[\vec S^2(x)-1\bigg]\\
 & \prodl_t\int\!\!\dm\vec m(t)\delta^{(N)}\bigg[\vec m(t)-\frac{1}{V_s}\int\!\!\dm\vec x\, \vec S(t,\vec x)\bigg] e^{-\Aeff(\vec S)}\,.
 \end{split}
\end{equation}
By choosing appropriate field redefinitions, which have been worked out in~\cite{Hasenfratz:2009mp}, $Z$ can be written as
\begin{equation}
 \begin{split}\label{eq: partition function in delta regime}
  Z & =  \prodl_t\int\!\!\dm\vec{e}(t)\,\prodl_x\int\!\!\dm\bpi(x)\,\delta^{(N-1)} \bigg[\frac{1}{V_s}\int\!\!\dm\vec x\,\bpi(t,\vec x)\bigg] e^{-\Aeff(\Omega\hat V^T\vec R)}
\end{split}
\end{equation}

It is indicated that in the effective action  the variable $\vec S(x)$ will be replaced by the combination
\begin{equation}\label{eq: S in terms of Omega and R}
\vec S(t,\vec x) =\Omega(t)\hat V^T(t)\vec R(t,\vec x)\,. 
\end{equation}
The $4$-component  vector $\vec R$ of unit length is parametrised as
\begin{equation}\label{eq: parametrisation of R in terms of pi}
  \vec R(x) = \left(\sqrt{1-\bpi^2(x)},\bpi(x) \right)\,.
\end{equation}
Eqs.~\eqref{eq: parametrisation of R in terms of pi} and~\eqref{eq: partition function in delta regime} ensure that the slow modes are not a part of the $\bpi$-fields. In fact the $k=(k_0,\vec k=\vec 0)$ modes have to be left out when we calculate the Green's functions. Furthermore, it follows that
\begin{equation}\label{eq: int over pi is zero}
\frac{1}{V_s}\int\!\!\dm \vec x\, \pi_i(t,\vec x)=0\,,\qquad i = 1,\ldots, N-1\,.
\end{equation}
$\Omega(t)$ and  $\hat V(t)$ are $\gr{O}{N}$ matrices\footnote{In
  paper~\cite{Hasenfratz:2009mp}, the notation $\Sigma (t)$ was introduced to
  our $\hat V(t)$. We decided for this change, since the quark condensate is
  denoted, here (and in many other works), by $\Sigma$.}. The first column of
the matrix $\Omega(t)$ is 
\begin{equation}\label{eq: define first column of Omega}
 \vec e_\alpha(t) = \Omega_{\alpha 0}(t)\,,\qquad \alpha=0,\ldots,N-1\,,
\end{equation}
and the matrix  $\hat V(t)$ has the following structure
\begin{equation}\label{eq: structure of the O(N) matrix V_hat}
\hat V(t) =\left( \begin{matrix}
  1& 0 \cdots 0  \\
  \begin{array}{c} 0 \\ \vdots\\  0\end{array} &  \bar V(t)
            \end{matrix}
\right) \,.
\end{equation}

The partition function~\eqref{eq: partition function in delta regime} is 
expressed as a path integral over the slow modes $\vec e(t)$ and as a path
integral over the fast modes $\bpi(x)$. The small
fluctuations (fast modes) can be integrated out in perturbation theory. Therefore, we have to expand the effective action  in terms of the $\bpi$-fields, up to the desired order. The
$\bpi$-fields enter by replacing the variables $\vec S$ according to the
parametrisations defined in Eq.~\eqref{eq: S in terms of Omega and R} and
Eq.~\eqref{eq: parametrisation of R in terms of pi}.

We have not defined the two matrices $\Omega(t)$ and $\hat V(t)$
completely. For  the symmetry-breaking terms the unknown
(undetermined) parts of $\Omega(t)$ and $\hat V(t)$ will drop out. Hence, the fixing of these two matrices will not be discussed in this work. A detailed
treatment of this issue is given in~\cite{Hasenfratz:2009mp}.


\section{The symmetry-breaking terms up to NNLO}\label{sec: expand
  symmetry breaking action up to NNL}

The terms~\eqref{eq: A20}-\eqref{eq: A02} of the effective action are still expressed in the field variable $\vec S$.  As mentioned before, we have to expand the effective action in Eq.~\eqref{eq:
  partition function in delta regime} in the $\bpi$-fields up to the desired
order (NNL). In this paper, the focus is set only on the symmetry-breaking terms, \ie the terms $\mathcal{A}_{(0,1)}$, $\mathcal{A}_{(2,1)}$ and $\mathcal{A}_{(0,2)}$. The expansion of the symmetric terms has been covered in~\cite{Hasenfratz:2009mp}.

\subsection{$\mathcal{A}_{(0,1)}$  up to NNLO}

The symmetry-breaking term of the $\ordo{p^2}$ effective action is given by 
\begin{equation}\label{eq: A_01 in terms of S}
 \mathcal{A}_{(0,1)}(\vec S) = -F^2M^2\int\!\!\dm x\,  S_0(x)\,.
\end{equation}
Using the relations~\eqref{eq: S in terms of Omega and R}, \eqref{eq: parametrisation of R in terms of pi}, 
\eqref{eq: define first column of Omega}, \eqref{eq: structure of the O(N) matrix V_hat} to express $\vec S$ in terms of $\bpi$-fields, $S_0$ can be written as
\begin{align*}
 S_0(x) & = \Omega_{0\alpha}(t)\hat V_{\beta\alpha}(t)R_\beta(x)	\\
	& = e_0(t)R_0(x) + \Omega_{0i}(t)\hat V_{ji}(t) \pi_j(x)
\end{align*}
We use the convention that Greek indices run from $0$ to $N-1$, whereas
Latin indices run from $1$ to $N-1$, and we will stick to this convention
throughout this paper, as long as nothing else is mentioned.

The term proportional to $\pi_j(x)$ contains some unknown elements of the matrices $\hat V$ and $\Omega$. According to~\eqref{eq: int over pi is zero} this term will, however, vanish when $S_0$ is plugged into Eq.~\eqref{eq: A_01 in
  terms of S}. After having expanded $R_0(x)$ in terms of the $\bpi$-fields, $\mathcal{A}_{(0,1)}$ reads as
\begin{equation}\label{eq: A_01 up to NNLO}
\mathcal{A}_{(0,1)} = -F^2M^2\int\!\!\dm x\, e_0(t)\left(1-\frac{1}{2}\bpi^2(x) +\ldots \right) \,.
\end{equation}
The dots indicate that there are terms of higher order in the
$\bpi$-fields. However, we can truncate the expansion already at this order,
since the leading contribution from the symmetry breaking will enter already
as a NLO correction in the rotator spectrum.

\subsection{$\mathcal{A}_{(0,2)}$ and $\mathcal{A}_{(2,1)}$ up to NNLO} 

The symmetry-breaking terms entering at $\ordo{p^4}$ are given by Eq.~\eqref{eq: A02}, and Eq.~\eqref{eq: A21} respectively. We will
give arguments why contributions from these two terms will only enter beyond NNLO. 

In order to get a crude estimate for the sizes of the corrections coming from
these two terms, we consider the system in a simplified form. We neglect the fast mode corrections. The variable $\vec S(x)$
can then be replaced by $\vec e(t)$, and the leading contribution from the symmetric
term simplifies to
\begin{equation}
 \mathcal{A}_{(2,0)} \stackrel{\bpi\to 0}{\approx} F^2L_s^3\int\!\!\dm t\,\frac{1}{2}\dot{\vec{e}}(t)\dot{\vec{e}}(t)
\end{equation}
which can be interpreted as the action of the $\gr{O}{4}$ rotator. The energy
eigenvalues of this system are proportional to $1/F^2L_s^3$. 

Under the same assumptions ($\bpi\to 0$), the symmetry breaking term $\mathcal{A}_{(0,2)}$ is reduced to the following  form
\begin{equation}
\mathcal{A}_{(0,2)} \approx -\int\!\!\dm t\, k_2 M^4 L_s^3 e_0(t)^2\,,
\end{equation}
This term can now be regarded as a perturbation to the symmetric rotator ($M$ is small).  The field $e_0$ is quantity of $\ordo{1}$, naively the spectrum of the rotator will receive a correction proportional to $\sim M^4L_s^3$
\begin{equation*}
\frac{1}{F^2L_s^3}+\ordo{M^4 L_s^3} = \frac{1}{F^2L_s^3}\left(1+\ordo{M^4L_s^6 F^2}\right)\,.
\end{equation*}
The relative correction to the unperturbed rotator energy is then proportional to $F^2L_s^6 M^5$. The latter combination, however, is of $\ordo{\delta^8}$. This can
 be shown easily by expressing the strength of the perturbation in terms of
 the dimensionless expansion parameters $\delN$ and $\delM$
\begin{equation*}
 F^2L_s^6M^4 = r^4\delta^6 = \ordo{\delta^8}\,.
\end{equation*}

Similar considerations lead to the same conclusion for the term $\mathcal{A}_{(2,1)}$.  Neglecting any fast modes corrections, the corresponding part of the action  reads as
\begin{equation}
 \mathcal{A}_{(2,1)} \approx\int\!\!\dm t\,\frac{F^2L_s^3}{2}\ededt 2 k_1  \frac{M^2}{F^2}  \,e_0(t)\,.
\end{equation}
This expression can be considered again as a small
perturbation to the unperturbed rotator spectrum. Assuming the field $e_0$, which is of $\ordo{1}$, to be a constant, the action reduces to the symmetric case.  The corrections to the moment of inertia is then proportional to $\frac{M^2}{F^2}$. The latter combination is of $\ordo{\delta^7}$, and therefore also these corrections are  beyond NNLO. In addition, we have completely neglected the fact that this term would not contribute in the first order of the expansion, due to odd
number of fields involved. Thus, the actual corrections would be even smaller than $\ordo{\delta^7}$.

\newpage


\section{Integrating out the fast modes}\label{sec: integrate out fast modes}

We consider again the partition function~\eqref{eq: partition function in delta
  regime} and insert the effective action, which has been expanded in terms of the fast modes, in the exponent $\exp(-\mathcal{A}_{\mathrm{eff}})$. While keeping the kinetic terms for $\vec e(t)$ and $\bpi(x)$ in the exponent, the remaining terms are expanded in a Taylor series and will be treated as interactions 
\begin{subequations}\label{eq: expanded partition function}
\begin{align}
 Z & = \prodl_t\int\!\!\dm\vec e(t)\,\prodl_x\int\!\!\dm\bpi(x)\,\prodl_{i=1}^{N-1}\delta\Big[\frac{1}{V_s}\int\limits_{\vec x}\pi_i(x)\Big] \nonumber \\
 & \phantom{=} \cdot \exp\bigg[-\intl_t\frac{F^2 L_s^3}{2}\ededt\bigg] \cdot \exp\bigg[-\intl_x\frac{F^2}{2}\cov{\mu}\bpi(x)\cov{\mu}\bpi(x)\bigg] \nonumber \\
\label{eq:Z expanded sy 0a}	& \phantom{=}\bigg\{1 + \int\limits_x \frac{F^2}{2} \ededt\bpi^2(x) +\ldots\bigg\}\\
\label{eq:Z expanded sy 0b}	& \phantom{=}\bigg\{1 - \int\limits_x \frac{F^2}{2}\left(-\frac{2}{\epsilon^2} Q_{ij}(t)\right)\pi_i(x)\pi_j(x)+\ldots\bigg\}\\	
\label{eq:Z expanded c}	& \phantom{=}\bigg\{1 +\frac{1}{2!}\int\limits_x\!\!\int\limits_y \frac{F^4}{\epsilon^2} Q_{i0}(t) Q_{j0}(t')\bigg[ \frac{1}{2}\bpi^2(x)\cov{0}\pi_i(x) - \Big(\bpi(x)\cov{0}\bpi(x)\Big)\pi_i(x)  \bigg] \nonumber \\
	& \phantom{\bigg\{1 + \frac{1}{2!}\int\limits_x\!\!\int\limits_y \left(\frac{F^2}{2}\right)^2} \cdot \bigg[ \frac{1}{2}\bpi^2(y)\cov{0}\pi_i(y) - \Big(\bpi(y)\cov{0}\bpi(y)\Big)\pi_i(y)  \bigg] +\ldots \bigg\}\\
\label{eq:Z expanded sy 0c}	& \phantom{=}\bigg\{1 - \int\limits_x \frac{F^2}{2}\Big(\bpi(x)\cov{\mu}\bpi(x)\Big)\Big(\bpi(x)\cov{\mu}\bpi(x)\Big) + \ldots\bigg\}\\
\label{eq:Z expanded sy g42}	& \phantom{=}\bigg\{1 -  g_4^{(2)}\int\limits_x \bigg[\frac{1}{2}\ededt\cov{\mu}\bpi(x)\cov{\mu}\bpi(x) + \nonumber\\
&\phantom{\bigg\{1-g_4^{(2)}\int\limits_x\bigg[} +\frac{1}{\epsilon^2} Q_{i0}(t) Q_{j0}(t)\cov{0}\pi_i(x)\cov{0}\pi_j(x)\bigg]    +\ldots \bigg\}\\
\label{eq:Z expanded sy g43}	& \phantom{=}\bigg\{1 -  g_4^{(3)}\int\limits_x \bigg[\frac{1}{2}\ededt\,\cov{0}\bpi(x)\cov{0}\bpi(x) + \frac{1}{\epsilon^2}Q_{i0}(t)Q_{j0}(t) \nonumber \\
&\phantom{\bigg\{1 +  \int\limits_x g_4^{(3)}}   \cdot\Big(\cov{0}\pi_i(x)\cov{0}\pi_j(x)+\frac{1}{2}\sum\limits_{k=1}^{3}\cov{k}\pi_i(x)\cov{k}\pi_j(x)\Big)  \bigg]  +\ldots  \bigg\}\\
%
%
%
\label{eq:Z expanded sb 0}	& \phantom{=}\bigg\{1 + \int\limits_ x F^2M^2 e_0(t)\bigg[1-\frac{1}{2}\bpi^2(x)+\ldots\bigg] \nonumber \\
	&\phantom{\Bigg\{ 1} + \frac{1}{2!}\int\limits_x\!\!\int\limits_y F^4M^4  e_0(t)e_0(t')\bigg[1-\frac{1}{2}\bpi^2(x) -\frac{1}{2}\bpi^2(y) \bigg]  + \ldots \bigg\}\,.
\end{align}%
\end{subequations}
At this point the fast modes can be integrated out in a systematic way, keeping only terms up to $\ordo{\delta^4}$. In Eq.~\eqref{eq: expanded partition function} we used the convention $x=(t,\vec x)$, and $y=(t',\vec y)$ respectively. Furthermore, it is assumed that after having integrated out the fast modes, the limit $\epsilon\to 0$ has to be taken. The symmetric terms~\eqref{eq:Z expanded sy 0a}~-~\eqref{eq:Z expanded sy g43} have been discussed in~\cite{Hasenfratz:2009mp}, where also the origin of the $\epsilon$ is explained in more detail. In the following, the focus is set on the additional symmetry-breaking part~\eqref{eq:Z expanded sb 0}.

The pairing of two $\bpi$-field components which live at the same space time point $x$  will give a contribution of the form
\begin{equation*}
 \langle \pi_i(x)\pi_j(x)\rangle \propto \frac{\delta_{ij}}{F^2}\Dstar{0}\,.
\end{equation*}
$\Dstar{0}$ denotes the finite volume propagator evaluated at $x=0$. The ${}^{*}$ indicates that the $k=(k_0,\vec k=0)$ modes have to be left out when calculating the propagator. In $d=4$ dimensions, the finite volume propagator is proportional to $L_s^{-2}$. The pairing  two $\bpi$-fields results in a contribution $\sim\delta^2$
\begin{equation}
 \frac{1}{F^2}\Dstar{0} = \frac{1}{F^2 L_s^2}\Gbars\,,
\end{equation}
where $\Gbars$, given in Eq.~\eqref{eq: value of gbarstar}, is the finite volume Green's function evaluated at $x=0$ in DR.

Let us consider expression~\eqref{eq:Z expanded sb 0} separately, that means we ignore cross terms with any of the symmetric interactions for now. Again, we integrate out the fast modes, and we keep only terms up to $\ordo{\delta^4}$. Hence, the partition function can be written as 
\begin{equation}\label{eq: Z sb expansion after fast mode pert}
\begin{split}
 Z &\propto \prodl_t \int\!\!\dm\vec e(t)\, e^{-\intl_t\frac{\Theta}{2}\ededt}\Bigg\{1+\int\!\!\dm t\,F^2 L_s^3 M^2  \bigg(1-\frac{3}{2}\frac{\Dstar{0}}{F^2}\bigg)e_0(t)\\
 & \phantom{\propto} +\frac{1}{2!}\int\!\!\dm t\int\!\!\dm t'\,(F^2 L_s^3 M^2)^2 \bigg(1-3\frac{\Dstar{0}}{F^2}\bigg)e_0(t)e_0(t')+\ldots\Bigg\}\,.
\end{split}
\end{equation}
where, up to NNL order $\Theta$ is given as quoted in Eq.~\eqref{eq: moment of inertia up to NNL}
The expression  inside the curly brackets in Eq.~\eqref{eq: Z sb expansion after fast mode pert}, can be written as an exponential again
\begin{equation}\label{eq: Z O(N) rotator incl SB}
 \begin{split}
  Z & \propto \prodl_t\int\!\!\dm\vec e(t)\, \exp\bigg[-\int\!\!\dm t\,\frac{\Theta}{2}\ededt-\eta e_0(t)\bigg]\,.
 \end{split}
\end{equation}
Here, we  introduced  the new expression $\eta$ which we have already given in Eq.~\eqref{eq: eta incl NNLO corrections} as
\begin{equation*}
\eta =F^2L_s^3M^2\bigg(1-\frac{3\Gbars}{2 F^2L_s^2}\bigg)\,.
\end{equation*}
Since the leading contribution  from the symmetry-breaking interactions is a $\ordo{\delta^2}$ correction to the rotator spectrum,  the corrections to $\eta$ are only considered up to $(FL_s)^{-2}$.

A similar argument can be used to neglect any cross terms between the symmetric interactions and the symmetry-breaking interactions (on the level of the fast modes). The only possible cross terms which would connect fast modes from the symmetric part with fast modes from the symmetry-breaking part will enter only beyond $\ordo{\delta^4}$.  Thus, the cross terms factorise in a simple way and we can write the symmetric part and the symmetry-breaking part as an exponential~\eqref{eq: Z O(N) rotator incl SB}.

The original problem, described by an effective field theory in $d=4$ Euclidean space-time dimensions, has been reduced to a 1-dimensional system described by the action
\begin{equation}\label{eq: O(N) rotator Lagrangian}
 \mathcal{A}= \int\!\!\dm t\,\frac{\Theta}{2}\ededt - \eta e_0(t)\,,
\end{equation}
where the  variables $\vec e(t)$ satisfy the constraint
\begin{equation*}
 \vec {e}(t)\vec{e}(t) =1\,.
\end{equation*}

In the chiral limit ($\eta=0$), we identify the system~\eqref{eq: O(N) rotator
  Lagrangian} as that of a symmetric $\gr{O}{4}$ quantum mechanical rotator, where up to NNLO, the spectrum is given by~\eqref{eq: O(4) rotator spectrum}. In the explicitly  broken case, the rotator is considered in an external \scar{magnetic} field which points along the zeroth direction.  $\eta$ is assumed to be small, so that the symmetry-breaking potential can be treated as a perturbation. The corrections to the unperturbed rotator   are then calculated in perturbation theory in quantum mechanics. Considering again corrections up to $\ordo{\delta^4}$, requires perturbation theory in quantum mechanics up to 4th order.


\section{The quantum mechanical $\gr{O}{4}$ rotator}\label{sec: quantum mechanical rotator}

In this section, we will do the discussion for general $N\geq 3$, although we are explicitly interested in  case  $N=4$, finally.  The system described by Eq.~\eqref{eq: O(N) rotator Lagrangian} can be identified as an $\gr{O}{N}$ rotator in a small external \scar{magnetic} field. The Hamilton operator for this constraint system is given by
\begin{equation}\label{eq: Hamilton operator of the full rotator}
\op{H} = \frac{1}{\Theta}\left(\frac{\op{L}^2}{2} - \lambda\op{e_0}\right)\,,
\end{equation}
where $\op{L}$ can be considered as the \scar{angular momentum}  operator in the $N$-dimensional internal space. Again, $\Theta$ is the moment of inertia and $\lambda$, given by
\begin{equation}\label{eq: define lambda in terms of eta and theta}
 \lambda = \eta\Theta\,,
\end{equation}
%
is the small, dimensionless parameter which denotes the strength of the external \scar{magnetic} field.

The energy gap of the system is defined as the
difference between the energy of the first excited state and the ground state energy. In the chiral
limit ($\lambda\to0$), the ground state energy is zero and the energy gap is simply
given by the energy of first excited state. Due to the presence of the
\scar{magnetic} field, the energy levels of the first excited state split up. For $N=4$ the first
excited state splits up into a singlet and a triplet. The triplet provides the lower energy difference and is identified as the energy gap. 
Up to~$\ordo{\lambda^2}$ the energy gap has already been
calculated in \cite{Leutwyler:1987ak}, but without taking into account fast
modes corrections.

 Besides taking into account also fast modes corrections, we are interested in corrections up to $\ordo{\lambda^4}$. Since $\lambda$ is assumed to be small, we write the energy spectrum of~\eqref{eq: Hamilton operator of the full rotator} as a power series in $\lambda$ up to $\ordo{\lambda^4}$. 
\begin{equation}\label{eq: energy of O(4) rotator including corrections}
 E_{\ell,k}(\lambda) =\frac{1}{\Theta}\left( \egy{0}{\ell,k} + \suml_{i=1}^{4}\lambda^i \egy{i}{\ell,k} +\ordo{\lambda^5}\right)\,.
\end{equation}
Here, $\egy{0}{\ell,k}$ denotes the spectrum in the chiral limit ($\lambda\to0$) which for general $N\geq 3$ is given by
\begin{equation}\label{eq: energy spectrum O(N) rotator}
 \egy{0}{\ell,k} = \egy{0}{\ell} = \frac{1}{2}\ell(\ell+N-2)\,,\qquad \ell=0,1,\ldots\,.
\end{equation}
In order to determine the coefficients $\egy{i}{\ell,k}$ we apply  standard perturbation theory in quantum mechanics.

The energy eigenstates of the unperturbed system can be represented by associated Jacobi polynomials
\begin{equation}
 \ket{\ell,k} \sim P_{\ell,k}^{(N)}(z)\,.
\end{equation}
The polynomials $P^{(N)}_{\ell,k}(z)$ are defined as
\begin{align}
 P^{(N)}_{\ell,0}(z) & = \frac{(-1)^\ell\Gamma\left(\frac{N-1}{2}\right)}{2^\ell\Gamma\left(\ell+\frac{N-1}{2}\right)}(1-z^2)^{-(N-3)/2}\left(\frac{\dm}{\dm z}\right)^\ell(1-z^2)^{\ell+(N-3)/2}\,,	\label{eq: jacobi poly eigenstates}	\\
 P^{(N)}_{\ell,k}(z) & =(1-z^2)^{k/2}\left(\frac{\dm}{\dm z}\right)^kP^{(N)}_{\ell,0}(z)\,, \label{eq: ass. jacobi poly eigenstates}
\end{align}
and have been normalised to $P_{\ell,k}^{(N)}(1)=1$. $k$ denotes the $\gr{O}{N-1}$ \scar{angular momentum} and runs from $=0,\ldots,\ell$. The variable $z$ takes values in the
interval $[-1,1]$ and can be identified with the direction of the \scar{magnetic} field in the internal $N$-dimensional space.  For fixed $N$ and $k$, two of these polynomials are orthogonal in the interval $[-1,1]$ with respect to the weighting function $w(z)=(1-z^2)^{(N-3)/2}$
\begin{equation}\label{eq: jacobi poly ortho}
\langle P_{\ell,k}^{(N)}(z)P_{\ell',k}^{(N)}(z)\rangle = \intl_{-1}^{1}\!\!\dm z\,(1-z^2)^{\frac{N-3}{2}}P_{\ell,k}^{(N)}(z)P_{\ell',k}^{(N)}(z)\propto \delta_{\ell\ell'}\,.
\end{equation}

In perturbation theory matrix elements of the following form have to be calculated repeatedly
\begin{equation}\label{eq: PT mat element}
 V^k_{\ell\ell'}\doteq \frac{\expv{\ell,k}{\op{z}}{\ell',k}}{\norm{\ell,k}} = \frac{\langle z P^{(N)}_{\ell,k}(z)P^{(N)}_{\ell',k}(z) \rangle}{\langle P^{(N)}_{\ell,k}(z)P^{(N)}_{\ell,k}(z)\rangle},
\end{equation}
where only matrix elements of the form $V^k_{\ell,\ell+1}$ for $k=0,\ldots,\ell$, and $V^k_{\ell,\ell-1}$ for $k=0,\ldots,\ell-1$ are nonzero. As a consequence, the  corrections to the orders $\ordo{\lambda}$ and $\ordo{\lambda^3}$ vanish
\begin{equation}
 \egy{1}{\ell,k}=\egy{3}{\ell,k}=0\,,
\end{equation}
and at a given order of $\lambda^{2n}$, $n=1,2,\dots$, only a finite number of terms survive the (in general) infinite sums for the coefficients $\egy{2}{}$, $\egy{4}{}$ respectively
\begin{align}
 \egy{2}{\ell,k} & =  \frac{V^k_{\ell \ell-1}V^k_{\ell-1 \ell}}{\egy{0}{\ell}-\egy{0}{\ell-1}} +   \frac{V^k_{\ell \ell+1}V^k_{\ell+1 \ell}}{\egy{0}{\ell}-\egy{0}{\ell+1}}\,, \label{eq: lambda2 coefficients}\\
\egy{4}{\ell,k} &=  \frac{V_{\ell\ell-1}^kV_{\ell-1\ell-2}^kV_{\ell-2\ell-1}^kV_{\ell-1\ell}^k}{(\egy{0}{\ell}-\egy{0}{\ell-1})^2(\egy{0}{\ell}-\egy{0}{\ell-2})} + \frac{V_{\ell\ell+1}^kV_{\ell+1\ell+2}^kV_{\ell+2\ell+1}^kV_{\ell+1\ell}^k}{(\egy{0}{\ell}-\egy{0}{\ell+1})^2(\egy{0}{\ell}-\egy{0}{\ell+2})} \nonumber\\
	& \phantom{=} - \egy{2}{\ell,k}\left[\frac{V^k_{\ell\ell-1}V^k_{\ell-1\ell}}{(\egy{0}{\ell}-\egy{0}{\ell-1})^2} + \frac{V^k_{\ell\ell+1}V^k_{\ell+1\ell}}{(\egy{0}{\ell}-\egy{0}{\ell+1})^2} \right]\,.\label{eq: lambda4 coefficients}
\end{align}
Using Eqs.~\eqref{eq: energy spectrum O(N) rotator}-\eqref{eq: lambda4 coefficients}, the energies $E_{\ell,k}$~\eqref{eq: energy of O(4) rotator including corrections} can be calculated for arbitrary $N\geq 3$. 
%
%


\section{The energy gap for $N=4$}

The energy gap for $N=4$ is defined as the difference between the energy of first excited state $(\ell=1,k=0,1)$ (singlet, triplet) and the ground state $(\ell=0,k=0)$. In Tab.~\ref{tab: lambda coefficients}, we quote the coefficients for the ground state energy correction and for the first excited state energy correction. The triplet ($\ell=1,k=1$) provides the lower energy difference, and therefore the energy gap is defined as
\begin{equation}
E_{L_s} = E_{1,1}(\lambda)-E_{0,0}(\lambda)\,.
\end{equation}
 Up to $\ordo{\lambda^4}$, the explicit result for energy gap reads as\footnote{The NL contribution is in agreement
  with~\cite{Leutwyler:1987ak}, as long as we neglect the fast mode
  corrections, \ie $E_{L_s} = 3/(2F^2L_s^3)\big[1+(F^8L_s^{12}M^4)/15+\ldots\big]$.}
\begin{equation}\label{eq:Egap rotator lambda^4}
 E_{L_s} = \frac{3}{2 \Theta}\left[1+\frac{\lambda^2}{15}-\frac{193}{120}\frac{\lambda^4}{15^2}\right]\,.
\end{equation}

Inserting the results for $\Theta$ and $\eta$, according to Eq.~\eqref{eq: moment of inertia up to NNL}, and Eq.~\eqref{eq: eta incl NNLO corrections} respectively, we recover the final result in Eq.~\eqref{eq: energy gap final result}. The leading term in $\lambda$ is given by the dimensionless expansion parameter $\delM^2$, and up to $\ordo{\delta^4}$ $\lambda$ reads as
\begin{align}
 \lambda^2 & =  F^8L^{12}M^4 \left[1 - \frac{7\Gbars}{(FL_s)^2}\right] +\ordo{\delta^6}\,.
\end{align}
\begin{table}[ht]
\centering
\begin{tabular}{lcccc}
\toprule
\multicolumn{2}{c}{$N=4$} &$\egy{0}{\ell,k}$ & $\egy{2}{\ell,k}$ & $\egy{4}{\ell,k}$\\
 \midrule
 $\ell =0$ &$k=0$ &  $0$ & $-\frac{1}{6}$ & $\phantom{-} \frac{5}{432}$ \\
 \\
 \multirow{2}{10mm}{$\ell=1$}&$k=0$ & $\frac{3}{2}  $  &	$\phantom{-}\frac{1}{15}	$ &			$-\frac{317}{27000}$					\\
 \\
 & $k=1$ & $\frac{3}{2}$ & $-\frac{1}{15}$ & $\phantom{-} \frac{23}{27000}$\\
 \bottomrule
\end{tabular}
\caption{Here, the coefficients $\egy{n}{\ell,k}$ ($n=0,2,4$) are given for the ground state ($\ell=k=0$), for the singlet $(\ell =1,k=0)$, and for the  triplet ($\ell=k=1$). }\label{tab: lambda coefficients}
\end{table}


\section{The constraints on  $L_s$ and $M$}
The formula for the energy gap Eq.~\eqref{eq: energy gap final result} involves two different expansions. The requirement that theses two expansion are applicable will put some constraints on the values of $L_s$ and $M$. These values can be considered as a rough estimate on the domain of $L_s$ and $M$, where the approximation used above is valid.

\begin{table}[ht]
\centering
\begin{tabular}{ c|cc|c|cc|cc}
\toprule
 $L_s[\fm]$		&  $\theta_\mathrm{NL}$	&  $ \theta_\mathrm{NNL}$  &  $\hat M_1[\MeV]$  & $\hat M_2[\MeV]$ &$\hat M_2 L_s$    \\
\midrule	
2.0					&			0.59	&	 -0.10		& 99		&		 213			&	2.16				\\
2.5					&		0.38		&		-0.05	&	80 	&	 	109	  		&	1.38			 \\
3.0					&		0.26		&		-0.03	&	66 	&  	63			&  0.96			 \\
\bottomrule	
\end{tabular}
\caption{This table shows the NLO ($\theta_{NL}$) and the NNLO ($\theta_{NNL}$) corrections to the moment of inertia for some selected values of $L_s$. $\hat M_0$ is simply $\hat M_1=1/L_s$, and $\hat M_2$ is defined according to~\eqref{eq: def hat M2}. As numerical input we used $F=86.2\MeV$, $\Lambda_1=120 \MeV$ and $\Lambda_2=1200\MeV$, obtained from~\cite{Colangelo:2003hf}.}\label{tab: corrections and limits on M}
\end{table}

\paragraph{What is the constraint on $L_s$?}
The moment of inertia $\Theta$ receives corrections proportional to $\delN^2$ at NLO, and proportional ot $\delN^4$ at NNLO respectively
\begin{equation}
\Theta = F^2L_s^3\left[1+\theta_\mathrm{NL}+\theta_\mathrm{NNL}\right]\,.
\end{equation}
In order to have this expansion working properly, the  corrections $\theta_{\mathrm{NL}}$ and $\theta_\mathrm{NNL}$ should be small. In fact, we require that $\theta_{\mathrm{NL}}$ should be roughly $50\%$ or smaller. This requirement puts a lower bound on $L_s$. An estimate for the lower bound of the box size can be obtained from Tab.~\ref{tab: corrections and limits on M}. There, the size of the corrections at NLO and at NNLO are given for some selected values of $L_s$. For $L_s=2.0\fm$ the NLO corrections turn out to be rather large, about $60\%$. Going to towards $L_s=2.5\fm$ the NLO corrections decrease to $40\%$. Thus, we conclude that volumes about $L_s\gtrsim 2.5\fm$  should be considered. 

\paragraph{What are the constraints on $M$?}
The $\delta$-regime requires the mass to be much smaller than the inverse box size. We define $\hat M_1$ to be
\begin{equation}
\hat M_1 = \frac{1}{L_s}\,.
\end{equation}
This relation gives a first, upper bound on the  mass for a given $L_s$. The corresponding values are quoted in Tab.~\ref{tab: corrections and limits on M}. Taking  $L_s=2.5\fm$ as a reference value for the lower bound on $L_s$,  the mass should be smaller than $M=80\MeV$.

 However, this constraint does not account for the fact that the expansion for the energy gap~\eqref{eq: energy gap final result} can brake down. This issue can be circumvented requiring the first correction to the energy gap ($r^4/15$) to be small, e.g.\ $50\%$ or smaller. We define $\hat M_2$ as the second upper bound satisfying the relation

\begin{equation}\label{eq: def hat M2}
  \frac{F^8L_s^{12}\hat M_2^4}{15}= \frac{1}{2}\,,
\end{equation}
for a given $L_s$. It turns out that this constraint only becomes relevant for larger values of $L_s$, \ie around $L_s=3\fm$.

From the considerations above we conclude that an appropriate choice for $L_s$ and $M$  is crucial, in order to be in the domain where the two expansions work properly, and the formula for the energy gap is valid.  We defined the lower bound on $L_s$ to be about $2.5\fm$. This forces the leading order term in the pion mass $M$ to be smaller than $80\MeV$, and therefore, we have to use quarks with masses below the physical quark masses. Since  $M$ scales roughly as $\sim 1/L_s^3$, the upper bound for the mass decreases even faster with increasing $L_s$.

\subsection*{Acknowledgements}
The author would like to thank P.~Hasenfratz, F.~Niedermayer, Ch.~Weiermann,
G.~Colangelo for the
helpful and interesting discussions. This work has been supported in part by
the Swiss National Science Foundation. The Albert Einstein Center for
Fundamental Physics is supported by the ``Innovations- und Kooperationsprojekt
C-13'' of the ``Schweizerische Universit\"atskonferenz SUK/CRUS''. The author
acknowledges support by DFG project ``SFB/TR-55''.

%

\end{document}